\documentclass{mem}
\usepackage{natbib}\usepackage{txfonts}\usepackage{balance}
\usepackage{graphicx}
\usepackage[a4paper]{hyperref}
\idline{81}{1}

\begin{document}

\title{
Observation and analysis of chromospheric magnetic fields
}
   
\subtitle{}

\author{
J. \,de la Cruz Rodr\'iguez\inst{1} 
\and H. \,Socas-Navarro\inst{2}, \\
M. \,van Noort\inst{1} 
\and L. \, Rouppe van der Voort\inst{3}
}

\offprints{J. de la Cruz Rodr\'iguez} 

\institute{
The Institute for Solar Physics of the Royal Swedish Academy of Sciences, 
AlbaNova, SE-106 91  Stockholm, Sweden, \email{jaime@astro.su.se}
\and
Instituto de Astrof\'isica de Canarias, Avda V\'ia L\'actea S/N, 
La Laguna 38205, Tenerife, Spain, \email{hsocas@ll.iac.es} 
\and
Institute of Theoretical Astrophysics, P.O. box 1029 Blindern, 
N-0315 Oslo, Norway, \email{v.d.v.l.rouppe@astro.uio.no}
}

\authorrunning{de la Cruz Rodr\'iguez et al. }

\titlerunning{Chromospheric magnetic fields}

\abstract{
The solar chromosphere is a vigorously dynamic 
region of the sun, where waves and magnetic fields play an important role. 
To improve chromospheric diagnostics, we present new 
observations in \ion{Ca}{II}~8542 carried out with the SST/CRISP on 
La Palma, working in full-Stokes mode. We measured Stokes line profiles 
in active regions. The line profiles observed 
close to the solar limb show signals in all four Stokes parameters, while 
profiles observed close to disk center only show signals above the 
noise level in Stokes I and V. We used the NLTE inversion code 'NICOLE' to 
derive atmospheric parameters in umbral flashes present in a small round 
sunspot without penumbra.
}

\maketitle{}

\section{Introduction}\label{introduction}

The magnetic structure of the solar chromosphere is currently the
subject of intense computational  and observational
research. Numerical MHD  simulations have  improved in the last
few years, becoming more and more realistic, i.e,
\cite{2004A&A...414.1121W}, \cite{2007A&A...473..625L} and
\cite{2009ApJ...701.1569M}. Unfortunately, progress on the
observational side has arguably been somewhat slower. The few optical
spectral lines probing the chromosphere are typically characterized
by:
\begin{enumerate}
\item A vast formation range, from the photosphere to the chromosphere.
\item NLTE.
\item A weak polarization signal (especially in Q and U).
\item A large spectral width.
\end{enumerate}

Inversion codes have been commonly used to derive atmospheric
parameters, usually assuming  LTE or considering a Milne-Eddington
atmosphere. However, this approximation cannot be used  for modeling
chromospheric lines, such as the Ca II infrared triplet. A method for
NLTE inversions of Zeeman-induced Stokes profiles was described by
\citet{2000ApJ...530..977S} and used in a study by
\citet{2007ApJ...670..885P} of slit-based spectroscopic observations.
The development of new instrumentation at large aperture telescopes
(like SST/CRISP, DST/IBIS, VTT/TESOS) has made it possible to observe
the chromosphere with very high spatial and temporal
resolution. Here, we present inversions of chromospheric full-Stokes
CRISP data.

\section{The chromosphere observed with CRISP in 854.2~nm}

CRISP is a Fabry-P\'erot filter consisting of two etalons mounted in
tandem. The rapid tuning ability in combination with two Liquid
Crystal Variable Retarders allow for  full-Stokes scanning of spectral
line profiles. The transmission profile of the instrument at 8542~\AA\ 
has  a FWHM of $\sim$~111~m\AA. The \ion{Ca}{II}~IR lines have a large
formation range: the wings of the line are formed in the photosphere,
while the inner core is formed in the lower chromosphere. Therefore,
the features observed in the core  of the line show significant
changes on time scales of 3--5~seconds, whereas the features in the
wings, showing the usual photospheric granulation pattern, evolve
much slower ($\sim$~25~seconds). As with any Fabry-Perot instrument,
there is a trade-off between cadence, sensitivity and number of
wavelengths. The datasets were processed using MOMFBD
\citep{2005SoPh..228..191V}. It is important to note that our
observations of active regions show similar chromospheric  structures
to those seen in H$\alpha$. 

\subsection{High resolution Umbral Flashes}\label{umbral}

\begin{figure*}[]
\resizebox{\hsize}{!}{
\includegraphics[clip=true]{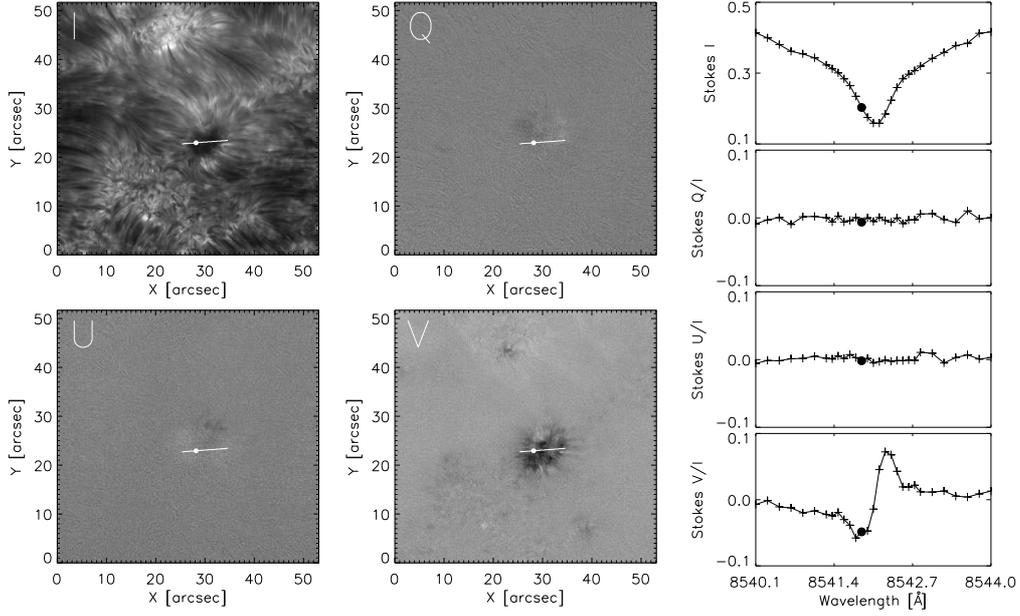}}
\caption{\footnotesize 
Observations at heliocentric angle $\mu=0.81$. 
Full-Stokes maps at $-182$~m\AA\ from LC in 854.2~nm (filled circle 
on the profiles panels). The line profiles plotted on the panels on the right 
correspond to the point marked with the circle in the images. }
\label{scan2}
\end{figure*}

In this study we present observations taken on 2008 June 12 at
$\mu=0.81$, with a cadence of 11 s per scan and 
covering a range of $\pm 1.94$~\AA\ from line center. 
The dataset, shown in Fig.~\ref{scan2},
includes a small sunspot without penumbra showing Umbral Flashes (UF,
hereafter), one  of the clearest examples of shock propagation in the
Sun. \citet{2000ApJ...544.1141S} report anomalous Zeeman profiles
during the UF phase and propose a scenario in which this phenomenon
may occur. \citet{2005ApJ...635..670C} provide further evidence for
the dual component scenario. \citet{2009ApJ...696.1683S} describe
fine-scale structures in UF from Hinode \ion{Ca}{II}~H imaging, further
supporting the idea of a highly structured umbral atmosphere. Our
observations show the anomalous Stokes V profile during the UF phase
in Fig.~\ref{wave}.
\begin{figure*}[]
\resizebox{\hsize}{!}{
\includegraphics[clip=true]{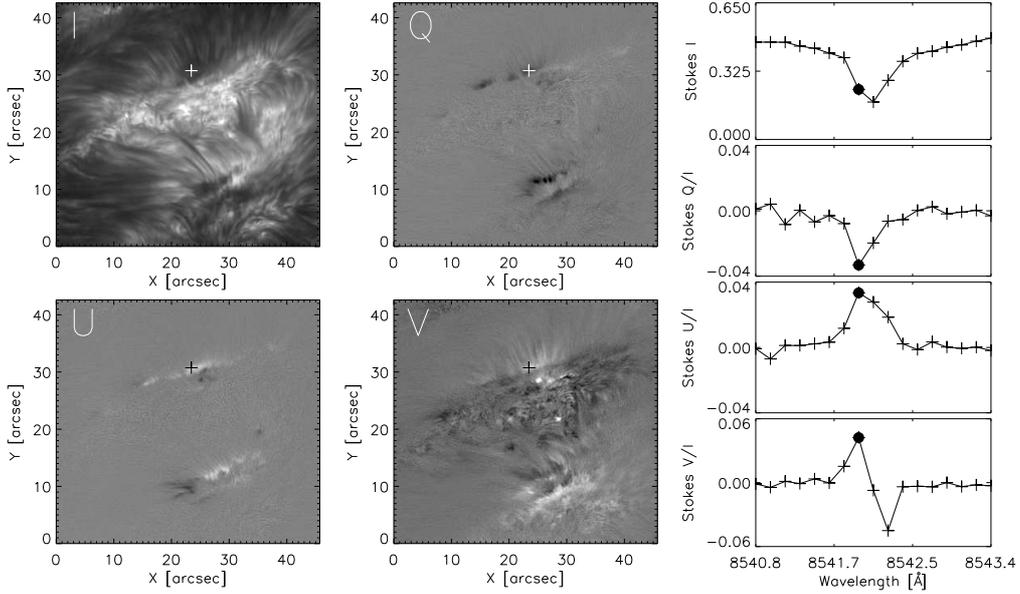}}
\caption{\footnotesize 
Observations at heliocentric angle $\mu=0.43$. 
Full-Stokes maps at $-161$~m\AA\ from line core in 854.2~nm (filled circle 
on the profiles panels). The line profiles plotted on the panels on the right 
correspond to the point marked with the cross in the images. }
\label{scan}
\end{figure*}

\subsection{Active region at $\mu=0.43$}

Figure~\ref{scan} shows an active region which encloses two small
sunspots recorded on 2008 June 10 at $\mu=0.43$. The Stokes images
clearly show a linear polarization signal, while the Stokes V image
suggests a very complicated magnetic field configuration. The
circular polarization peaks around $\pm 6$\%, while the linear
polarization measured in Stokes Q \& U is (in most cases) lower than 
4\%. No inversions of this dataset were carried out. 

\section{NICOLE: the NLTE inversion code}

We use the NLTE inversion code NICOLE, which assumes complete
redistribution in frequencies and angles, and works with
plane-parallel geometry. Line broadening produced by collisions
with neutral H atoms follows the formulation described by
\citet{2000A&AS..142..467B}. The code is based on the work presented
in \citet{2000ApJ...530..977S}. It solves the NLTE problem using the
strategy described by \citet{1997ApJ...490..383S}

The inversions are initialized with a modified HSRA atmospheric model
and the code uses CRISP's non-Gaussian instrumental profile to degrade
the synthetic  spectra. Similar NLTE inversions are described in
detail by \citet{2007ApJ...670..885P}, but concerned quiet Sun
and at a much lower spatial resolution.

We use NICOLE to invert the observations described in Sect.~\ref{umbral},
with the aim of testing the code on CRISP observations for further
analysis of the results. The weak Stokes signals, high image
resolution, limited wavelength coverage and poor sampling of the line
present a challenge for the analysis of such data. We carry out the
inversions using only a single component atmosphere, even though UF
probably require a multi-component treatment, and therefore, the results
represent an average of multiple components. Figure~\ref{scan2} shows
the selected target for the inversions, where the circle on the image
marks an example pixel of which the inversion results  are shown in Sect.~\ref{res}.

\section{Results}\label{res}

\begin{figure}[]
\resizebox{\hsize}{!}{
\includegraphics[clip=true]{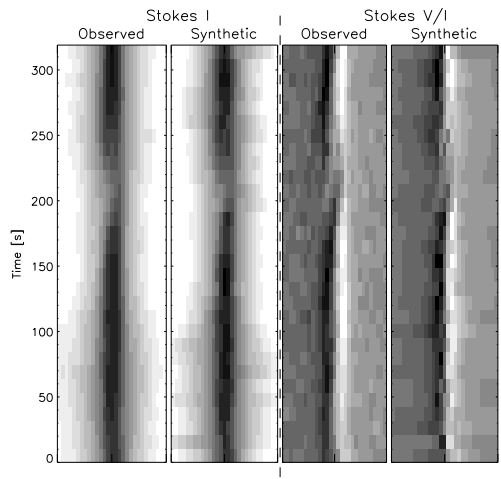}}
\caption{\footnotesize 
Time-series of the observed and synthetic spectra. The pixel used 
in this example is marked with a circle in Fig.~\ref{scan2}.}
\label{panels}
\end{figure}

Unfortunately, NLTE inversions are computationally very expensive,
making the analysis of time series of  the whole field of view a time
consuming task. Therefore, we have inverted spectra only along a line
that crosses the pore for a 5~minutes time series. Figure~\ref{panels}
shows the time evolution of the spectrum of our example pixel. The UF
starts around $t=180$~s and ends at $t=220$~s.
\begin{figure}[]
\resizebox{\hsize}{!}{
\includegraphics[clip=true]{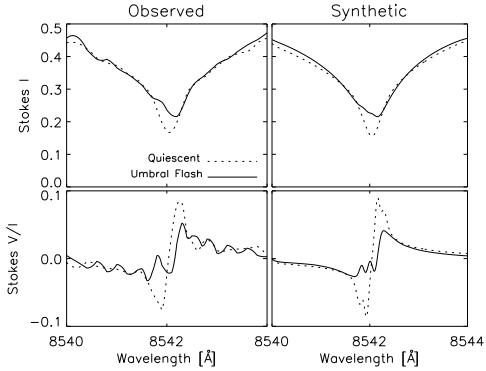}}
\caption{\footnotesize 
Observed and synthetic profiles in the 
quiescent (dashed line) and UF (solid line) phases. The pixel used 
in this example is marked with a circle in Fig.~\ref{scan2}.  }
\label{wave}
\end{figure}

\begin{figure}[]
\resizebox{\hsize}{!}{
\includegraphics[clip=true]{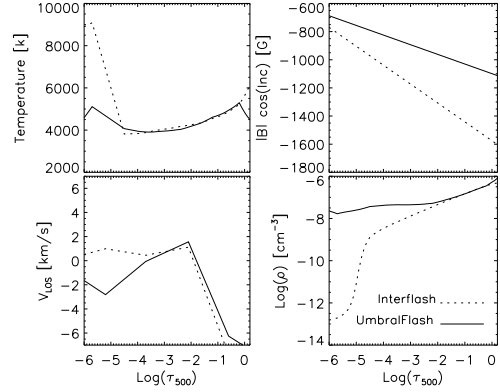}}
\caption{\footnotesize 
Model atmospheres obtained by inversions of Stokes 
profiles in Fig.~\ref{wave}. }
\label{model}
\end{figure}

In the quiescent phase our target shows a traditional sunspot Stokes
profile, with two $\sigma$ Zeeman components in circular polarization
with opposite signs. In the presence of an UF, the Stokes I profile
shows an enhanced emission component on the blue side of the line
core, which causes an apparent redshift of the line core, presented
in Fig.~\ref{wave}. The shock has a strong signature in the Stokes V
profile and  makes the blue Zeeman $\sigma$ component
vanish. Figure~\ref{model} shows the atmospheric model obtained by
inversions of the Stokes profiles of the example pixel during the
quiescent (dashed line) and UF (solid line) phases. 
The UF is characterized by an upflow (negative velocity) in the higher
layers of the model, compared to the quiescent phase.
The density appears enhanced
during the UF phase, and the measured magnetic field strength is lower
due to the  presence of the anomalous Stokes V profile. The azimuth
and inclination of the magnetic field  could not be derived from the
Stokes data since the signal level in Stokes Q and U is too low. The
round shape of  the pore and the absence of a penumbra suggest that
the magnetic field is mostly vertical, which for high $\mu$ values is
roughly aligned with the line of sight, consistent with the low signal
in Stokes Q and U.

\section{Conclusions}

We have reduced polarimetric CRISP observations of the \ion{Ca}{II}~854.2~nm
line and analyzed a small, round pore located at $\mu=0.81$ using an
adapted version of the NLTE inversion code NICOLE. The code is
not able to reproduce discontinuities  as the model is smoothed with a
spline fit during the inversion. Nevertheless, the shock leaves a
clear fingerprint in the derived velocity, density and magnetic field
strength. Since the observed Stokes V profile of this line is known to
be properly described by Zeeman splitting in such a strong magnetic
field, the spread in the magnetic field strength is probably mostly
caused by noise. The amount of cross-talk between other thermodynamic
parameters  is still unknown and will be the subject of future
research.

\begin{acknowledgements}
This research project has been supported by a Marie Curie Early Stage Research
Training Fellowship of the European Community's Sixth Framework Program under
contract number MEST-CT-2005-020395: The USO-SP International 
School for Solar Physics.
\end{acknowledgements}

\bibliographystyle{aa}
\bibliography{delacruz}

\begin{thebibliography}{11}
\expandafter\ifx\csname natexlab\endcsname\relax\def\natexlab#1{#1}\fi

\bibitem[{{Barklem} {et~al.}(2000){Barklem}, {Piskunov}, \&
  {O'Mara}}]{2000A&AS..142..467B}
{Barklem}, P.~S., {Piskunov}, N., \& {O'Mara}, B.~J. 2000, \aaps, 142, 467

\bibitem[{{Centeno} {et~al.}(2005){Centeno}, {Socas-Navarro}, {Collados}, \&
  {Trujillo Bueno}}]{2005ApJ...635..670C}
{Centeno}, R., {Socas-Navarro}, H., {Collados}, M., \& {Trujillo Bueno}, J.
  2005, \apj, 635, 670

\bibitem[{{Leenaarts} {et~al.}(2007){Leenaarts}, {Carlsson}, {Hansteen}, \&
  {Rutten}}]{2007A&A...473..625L}
{Leenaarts}, J., {Carlsson}, M., {Hansteen}, V., \& {Rutten}, R.~J. 2007, \aap,
  473, 625

\bibitem[{{Mart{\'{\i}}nez-Sykora} {et~al.}(2009){Mart{\'{\i}}nez-Sykora},
  {Hansteen}, {DePontieu}, \& {Carlsson}}]{2009ApJ...701.1569M}
{Mart{\'{\i}}nez-Sykora}, J., {Hansteen}, V., {DePontieu}, B., \& {Carlsson},
  M. 2009, \apj, 701, 1569

\bibitem[{{Pietarila} {et~al.}(2007){Pietarila}, {Socas-Navarro}, \&
  {Bogdan}}]{2007ApJ...670..885P}
{Pietarila}, A., {Socas-Navarro}, H., \& {Bogdan}, T. 2007, \apj, 670, 885

\bibitem[{{Socas-Navarro} {et~al.}(2009){Socas-Navarro}, {McIntosh}, {Centeno},
  {de Wijn}, \& {Lites}}]{2009ApJ...696.1683S}
{Socas-Navarro}, H., {McIntosh}, S.~W., {Centeno}, R., {de Wijn}, A.~G., \&
  {Lites}, B.~W. 2009, \apj, 696, 1683

\bibitem[{{Socas-Navarro} \& {Trujillo Bueno}(1997)}]{1997ApJ...490..383S}
{Socas-Navarro}, H. \& {Trujillo Bueno}, J. 1997, \apj, 490, 383

\bibitem[{{Socas-Navarro} {et~al.}(2000{\natexlab{a}}){Socas-Navarro},
  {Trujillo Bueno}, \& {Ruiz Cobo}}]{2000ApJ...544.1141S}
{Socas-Navarro}, H., {Trujillo Bueno}, J., \& {Ruiz Cobo}, B.
  2000{\natexlab{a}}, \apj, 544, 1141

\bibitem[{{Socas-Navarro} {et~al.}(2000{\natexlab{b}}){Socas-Navarro},
  {Trujillo Bueno}, \& {Ruiz Cobo}}]{2000ApJ...530..977S}
{Socas-Navarro}, H., {Trujillo Bueno}, J., \& {Ruiz Cobo}, B.
  2000{\natexlab{b}}, \apj, 530, 977

\bibitem[{{van Noort} {et~al.}(2005){van Noort}, {Rouppe van der Voort}, \&
  {L{\"o}fdahl}}]{2005SoPh..228..191V}
{van Noort}, M., {Rouppe van der Voort}, L., \& {L{\"o}fdahl}, M.~G. 2005,
  \solphys, 228, 191

\bibitem[{{Wedemeyer} {et~al.}(2004){Wedemeyer}, {Freytag}, {Steffen},
  {Ludwig}, \& {Holweger}}]{2004A&A...414.1121W}
{Wedemeyer}, S., {Freytag}, B., {Steffen}, M., {Ludwig}, H., \& {Holweger}, H.
  2004, \aap, 414, 1121

\end{thebibliography}

\end{document}